# A Question Answering System Using Graph-Pattern Association Rules (QAGPAR) On YAGO Knowledge Base

*Wahyudi[1,2], Masayu Leylia Khodra[2], Ary Setijadi Prihatmanto[2], Carmadi Machbub[2]*

[1] [1)]Faculty of Information Technology, University of Andalas
Padang, Indonesia
[2)]School of Electrical Engineering and Informatics, Institut Teknologi Bandung
Bandung, Indonesia

wahyudi@fti.unand.ac.id, masayu@stei.itb.ac.id, asetijadi@lskk.ee.itb.ac.id, carmadi@stei.itb.ac.id

**Abstract.** A question answering system (QA System) was developed that uses graph-pattern association rules on the YAGO knowledge base. The answer as output of the system is provided based on a user question as input. If the answer is missing or unavailable in the database, then graph-pattern association rules are used to get the answer. The architecture of this question answering system is as follows: question classification, graph component generation, query generation, and query processing. The question answering system uses association graph patterns in a waterfall model. In this paper, the architecture of the system is described, specifically discussing its reasoning and performance capabilities. The results of this research is that rules with high confidence and correct logic produce correct answers, and vice versa.

## 1. Introduction

Knowledge bases provide information about a large variety of entities, such as people, countries, rivers, etc. Moreover, knowledge bases also contain facts related to these entities, e.g., who lives where, which river is located in which city [1]. The YAGO knowledge base [2] is a commonsense knowledge base, i.e. a collection of commonly known facts and information. YAGO was built from extraction of Wikipedia using the WordNet ontology. However, WordNet's ontology is limited, so YAGO developed their own ontology. The original YAGO evolved into YAGO2 [3] with the addition of data extracted from GeoNames, so it can describe spatial entities using spatial data from GeoNames, such as longitude and latitude. The expansion of YAGO2 into YAGO3 [4] added multilinguality; in previous versions, data were only extracted from English Wikipedia. YAGO3 also groups entities based on languages supported by Wikipedia.

Knowledge-based question answering systems can answer questions that require specialized knowledge, where much information needs to be summarized and incorporated in real time rather than just having a paragraph of text that was retrieved previously. In ontology-based question answering systems [5], the knowledge base where the answers are sought, has a structured organization defined by the ontology. Users can ask a question in natural language and the system will return accurate answers directly after analyzing the question, retrieving information and extracting the answer. Ontology-based knowledge bases provide a convenient way to gain knowledge for users, but natural languages need to be mapped into queries. Previous research has shown that there are two major challenges in ontology-based question answering systems that need to be addressed. Firstly, to understand the intent of the user in the question analysis stage and formally represent it. Secondly, to translate these formal representations into the correct queries adapted to a formal basic ontology request scheme. The first problem can be solved through natural language processing technology, while for the second question it is very important to look closely at the ontology to get answers.

The question answering system can only provide answers based on an existing dataset. The system cannot provide an answer if it is not available in the dataset. We attempted to solve this problem by using graph-pattern association rules, which allow the system to provide predictions from the rule patterns in the dataset [6].

In this paper, we describe a question answering system using graph-pattern association rules on the YAGO knowledge base. The rest of this paper is organized as follows: related work is discussed in the next section (Section 2), after which graph-pattern association rules are described in Section 3. In Section 4, we present the system architecture of the proposed question answering system using graph-pattern association rules on the YAGO knowledge base. An evaluation of the system's performance will be given in Section 5. Finally, Section 6 contains the conclusions of this paper.

## 2. RELATED WORK

A question answering system (QA system) is the process of finding the right answer to natural language questions. Such a system was first introduced in the early 1970s. It was used for problem solving in a particular domain. In QA systems, three main approaches are used: 1. natural language processing (NLP) systems, which map user questions into formal form, ensuring to get the most relevant answers; 2. information retrieval systems, used in conjunction with natural language processing, which focus on the extraction of facts from a number of large texts; 3. template-based systems, which match questions from users with a number of templates that have been prepared to get the information needed to produce the precise answer[7].

We focused on the third approach for the following reasons. A template-based QA system extends the natural language pattern matching approach to a database. System intelligence is manifested by a collection of manually created question templates. The first QA system that used this model was START. The START system makes use of annotations of knowledge for use in analysis. These annotations consist of: *subject-relation-object*. START is suitable for lexical units and sentence structures. START processes user questions to match them to annotations and returns answers based on the information obtained from the annotations. START has a virtual database called Omnibase. Omnibase consists of semi-structured data such as the *CIA Fact Book* and the Internet Movie DataBase (IMDB). The data model used in Omnibase is: *object-property-value*. Seiders' research also used template-based questioning systems. This system uses a common question template to get relationships between knowledge domains. The system uses 24 template types. Template-based QA systems are rarely studied by academics but are successfully applied in commercial applications, such as www.ask.com, that use the START approach [7].

Adolph et al. [6] created a QA system using the YAGO KB as the dataset. It uses RDF triple data and SPARQL queries. In general, the system consists of three components namely: 1. YAGO as repository to answer questions in RDF triple form via SPARQL; 2. A repository of text patterns and paraphrased questions compiled with two approaches: automatic processing using ClueWeb to obtain patterns connected to relationships in YAGO and collecting paraphrases for questions by crowdsourcing to humans; 3. question processing to decompose natural language questions into the form of a RDF triple pattern in SPARQL.

Moussa et al. [8] developed a QA system using the YAGO Kb that can be used to recognize questions in the form of 5W 1 H. The system was developed further to be able to convert natural language questions into RDF triple form.

Unger et al. [9] created a QA system using RDF datasets, one of which is YAGO. They used SPARQL question templates to convert the questions into RDF triple form. Their system is more about query optimization to answer questions that cannot be solved by regular or RDF triple queries. For example: What cities have more than 3 campuses? Or, mention 3 cities in Indonesia. This system creates a query pattern for such questions, which cannot be isolated with the usual RDF triple relationships. This system uses the SPARQL database.

The approach used in the above systems is different from the approach we used. Our research used graph property[10] database using neo4j as the graph data base and graph processing[11]. In neo4j graph processing using cypher query. In addition we can also provide answers even though the answer is not available in the data set by using graph-pattern association rule approach.

## 3. QAGPAR SYSTEM ARCHITECTURE

The architecture of the proposed QA system using graph-pattern association rules (QAGPAR) on the YAGO knowledge base is as follows: question classification, graph component generation, query generation, and query processing. The system uses graph-patterns in a waterfall model. Input questions in natural language are translated into the form of a graph. This is done for two important reasons: 1) the dataset uses graph properties; 2) by adopting a graph model, the searching process on the dataset will be facilitated and speeded up. Next, the query is determined based on the graph model. If the answer does not exist, then it will be processed using graph-pattern association rules. The answer to be given in accordance with the query will be translated from the graph pattern to natural language and given to the user, as shown in Figure. 1 below.

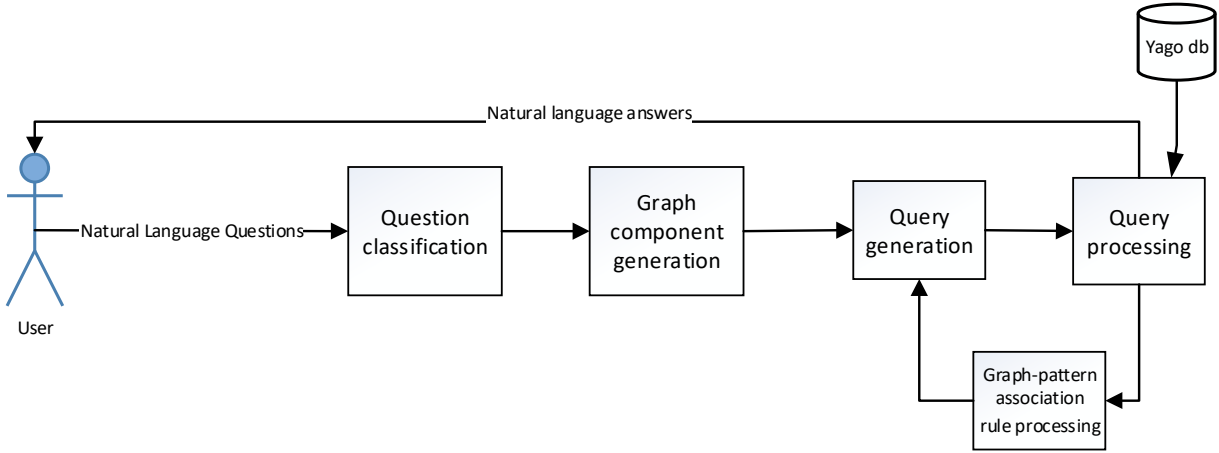

Figure 1. GAGPAR architecture

Generally, the whole process of system is divided into three major phases:

**Question analysis.** This phase processes natural language requests to obtain a formal graph representation to match the graph database. This requires several stages, such as classifying and deciphering the question. The output of the analysis is the query pattern, i.e. natural language requests are labeled with morphological information and ontology concepts.

**Query processing.** A given graph-shaped query pattern will be processed and executed to get the answer in node or edge form. If no answer is obtained the association graph pattern will be used to get the desired edges or nodes.

**Getting the answer.** This process is done to convert the nodes and edges that have been obtained into natural language and return the result to the user.

In this research we used three classes of patterns, with 37 patterns in total. Pattern I consists of 18 patterns, Pattern II consists of 18 patterns, while Pattern III consists of 1 pattern.

## A. Question classification

Question classification matches the question according to the question answering system. We use the question template technique, which means that the question in natural language that is entered must match an existing template. If the question does not match any template, then the question will not be processed. The output of this section is parsing the question. The output we expect in this process is different for each pattern. In Pattern I and Pattern II, the output that is captured contains the subject and the relation to the question. The relation that is used is different for each class of patterns. Some examples of templates used are given in Table I.

Table 1. Question template for question classification

| Pattern | Input | Output |
|---------|-------|--------|
| I | Who did (*p) marry? | (*p) and isMarriedTo |
| II | When was (*p) died? | (*p) and diedOnDate |
| III | What is the relationship between (* p) and (* p1) | (*p) and (*p1) |

## B. Graph component generation

The QAGPAR system will perform the transformation of the output at the stage of question classification into graph form. For Pattern I, we get a graph component with a node and an edge, for Pattern II we get a node with a property on that node, for Pattern III we get only two nodes. This is done so that the system can interpret the query to be used correctly. The transformation is done as shown in Table II.

Table 2. graph components used

| Pattern | input | output |
|---|---|---|
| I | (*p) and isMarriedTo | (*p) as node name and isMarriedTo as edge label 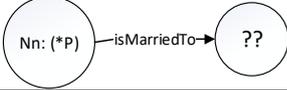 |
| II | (*p) and diedOnDate | (*p) as node name and diedOnDate as node property (*p) 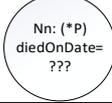 |
| III | (*p) and (*p1) | (*p) as node name 1 and (*p1) as node name 2 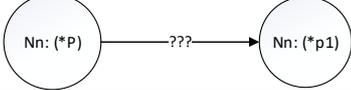 |

## C. Query generation

This stage generates queries based on applying the output model to the graph components. The complete process is shown in Table III.

Table 3. Query generation

| Pattern | input | Output |
|---|---|---|
| I | (*p) as node name and isMarriedTo as edge label | **'MATCH (n:owl_Thing{nn:'(*p)'})-[r{rdf_type:'isMarriedTo'}]-> (o) RETURN o '** |
| II | (*p) as node name and diedOnDate as node property (*p) | **'match(n:owl_Thing{nn:'(*p)'}) WITH n.diedOnDate as v return v'** |
| III | (*p) as node name 1 dan (*p1) as node name 2 | **'MATCH (n:owl_Thing{nn:'(*p)'})-[r]->(o:owl_Thing{nn:'(*p1)'}) RETURN r'** |

## D. Query processing

The next process is query execution to get the answer from the database. If the answer is not obtained, then the system will provide a notification that the answer does not exist. In contrast to the other patterns, for Pattern I, when the answer does not exist in the database, the system will optimize the query using graph-pattern association rules. For example: for Pattern I in Table 2, the following graph-pattern association rule is used:

1. (a)-[hasChild]->(b)<-[hasChild]-(d)=> (a)[isMarriedTo]->(d)

2. (a)-[actedIn]->(b)<-[directed]-(d)=> (a)[isMarriedTo]>(d)

3. (a)-[diedIn]->(b)<-[wasBornIn]-(c)<-[hasChild]-(d)   => (a)-[isMarriedTo]->(d)

In natural language, the rule is:

1. If A has child B and D has child B, then A is married to D

2. If A acted in film B and D directed film B, then A is married to D

3. If A died in city B, C was born in city B, and D had child C, then A is married to D

Table 4. Standard vs. PCA confidence rules

| Rules | Standard confidence | PCA confidence |
|-------|---------------------|----------------|
| 1 | 0.5859 | 0.7164 |
| 2 | 0.0036 | 0.0191 |
| 3 | 0.0005 | 0.0009 |

The three association rules each have a level of confidence. In the previous research we used 2 kinds of confidence measures, i.e. standard confidence and PCA confidence. Table 4 above gives the values of standard confidence and PCA confidence of the rules.

## E. System Algorithm

QAGPAR has two system checkbox interfaces, namely a normal and a graph-pattern association rule checkbox interface. The algorithm used is shown in Figure. 2.

```
Input : a question (S)
Output: an answer (A(b))
    1. AnalyzeQuestion(S, type,rel, v1, v2, prop)
    2. LoadYago(rel,v1,v2,n,prop,collection)
    3. if type = I and rel ≠ null and v1≠ null
        a. find v2 for rel (v1,v2)
        b. if v2≠ null set b = v2
        c. else (if v2 = null)do
                i. running for query-association(rel,v1)
               ii. loadYago(rel,v1,v2,prop,collection)
              iii. find v2 that rel(v1,v2)
               iv. if v2≠ null set b = v2
                v. else return "jawaban tidak ada"
    4. else if type=II and v1≠ null and prop≠ null
        a. find n for (v1,v1.prop = n)
        b. if n≠ null, set b = n
        c. else return "jawaban tidak ada"
    5. else if type =III and v1≠ null and v2≠ null
        a. find rel for rel(vj,vi)
        b. if rel≠ null set b = rel
        c. else return"jawaban tidak ada"
    6. return A(b)
```

Figure 2. QAGPAR system algorithm

The algorithm has an input question and the output is an answer. The input question is analyzed based on the question, the kind of pattern, relation (rail), node 1 (v1), node 2 (v2), and property (prop). Questions with Pattern I have relation and node 1, while node 2 and property are empty. Pattern II has node 1 and property node 1, while node 2 and relation are empty. Pattern III has node 1 and node 2, while relation and properties are empty. When the class of the question is known, then the next step is to find the required part. For Pattern I questions, find v2 so as to form the node and edge (v1, v2) pairs. For Pattern II, search the relation so as to form pairs of nodes and edges (v1, v2). For Pattern III, look for the value of node property v1. If any pattern finds an answer, the system will give the answer, but if no answer is found (node, edge or property value is not found), then the system will give no answer. Different from the other patterns, if no answer is found to a Pattern-I question, then we can predict an answer using graph-pattern association rules. For example: to the question 'Who is married to Malekeh_Jahan?'(Siapa pasangan dari malekeh_Jahan?), the system did not find an answer using normal search (Figure. 3 (a)). However, when trying graph-pattern association rules, the system can predict an answer based on the association rules made. In this case, the answer is 'Malekeh Jahan is married to Shah Qajar'(Malekeh Jahan adalah pasangan Shah Qajar) (Figure. 3(b)).

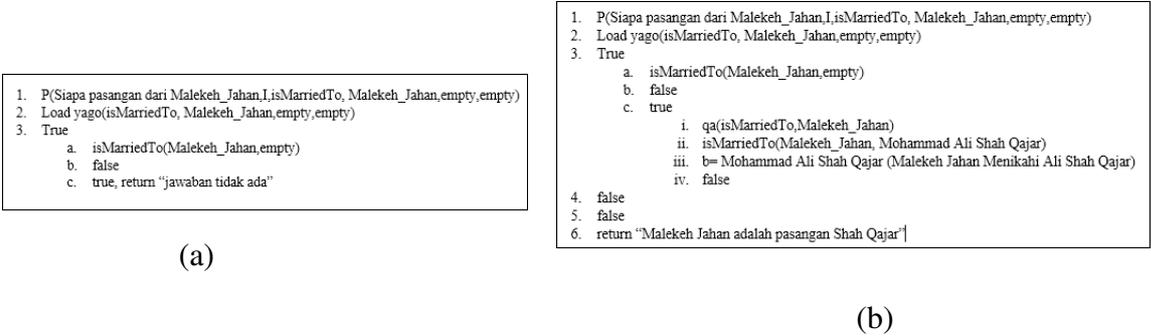

(a)

(b)

Figure 3. Example of QAGPAR algorithm

# 4. Graph pattern association rule

We use graph properties to mine association rules [10] for the following reasons: we follow Agrawal's [12] research model by starting from 1 antecedent and increasing the antecedents to 2, 3 and 4. With the tools we use [11], we can automatically search for the graph pattern of a query. However, there is a problem, i.e. there is only one edge direction. This is certainly different from the actual data in the database: the relationship can come from two sides of the subject or object.

In this section, we will explain the confidence values that we used, i.e. standard confidence and PCA confidence, which we adopted from Luis [13] and Fan [14].

**Confidence.** This is a measure to determine the strength of a rule. The value is between 0 and 1. A rule that has high confidence is close to 1 and vice versa.

**Standard Confidence**. Standard confidence (conf) is a measure of the ratio of the number of rules, $R$, compared to the facts we know in the form of graph pattern $P(G)$, as in the equation (1) below:

$$conf(R_i) = \frac{|R_i|}{|P_{(G)i}|} \qquad (1)$$

**PCA Confidence**. Standard confidence does not distinguish between things that are not in the dataset and wrong things that are in the dataset. In other words, standard confidence cannot distinguish between wrong facts and unknown facts. Since the knowledge base has no negative facts, we used the partial completeness approach. Luis [15] proposed to used partial completeness assumption (PCA). If $r(u, w) \in G$ for nodes $u$ and $w$, then:

$$\forall w' = r(u, w) \in G \cup new\ true \Rightarrow r(u, w') \in G$$

In other words, we assume that if graph $G$ knows some $x$-attribute of $u$, then all $x$-attributes of $u$ can be seen. This assumption is converted to standard confidence, obtaining:

$$PCA\ conf(R_i) = \frac{|R_i|}{|P(G)_i \wedge r(u,w')|} \qquad (2)$$

# 5. Experimental results

In the implementation of QAGPAR, YAGO ontology version 2017-W40-2 was utilized and converted to a Neo4J graph database. Java was used for developing the system. QAGPAR can provide answers according to the pattern input used, but when using graph-pattern association rules there are some issues, which will be discussed below.

We put the question: 'Who is married to Kurt_Brändle?', the system searched for the answer in the available database. Apparently, the answer was not in the database and the system gave as output 'no answer'. However, by using association rules, the same question was asked again to the system, and by using the rule 'Kurt_Brändle died in city C, B was born in city C and Julius van Zuylen van Nijevelt had child B, then Kurt Brändle is married to Julius van Zuylen van Nijevelt'. This rule has a low confidence level of 0.0005 standard confidence and 0.0054 PCA confidence. From the logic of the answer given it is concluded that the answer must be false, since Kurt_Brändle lived from 1912 to 1943 and was male, while Julius van Zuylen van Nijevelt lived from 1819 to 1868 and was also male, see Figure. 4(a).

Figure. 4(b) has the question 'Who is married to Jeremy Piven?' Using normal search, no answer was found in the database, so the system gave no answer. When using graph-pattern association rule, the question will be linked to the association rules. 'If Jeremy Piven plays in film B and Scott Marshall (director) is the director of film B then Jeremy Piven is married to Scott Marshall (director).' This rule has a confidence level of 0.0036 standard confidence and 0.0191 PCA confidence. From the logic of the answer it is concluded that the answer given is false because both Jeremy Piven and Scott Marshall (director) are male, although in terms of time there is no problem.

Figure 4(c) explains the question: 'Who is married to Malekeh_Jahan?' By using a normal search, no answer was found in the database, so the system gave no answer. When using the graph-pattern association rule 'If Malekeh_Jahan has a son, B, and Mohammad Ali Shah Qajar has a son, B, then Malekeh_Jahan is married to Mohammad Ali Shah Qajar.' This rule has a confidence level of 0.5859 standard confidence and 0.7164 PCA confidence. From the logic of the answer, it can be concluded that the answer 'Malekeh_Jahan is married to Mohammad Ali Shah Qajar' is correct.

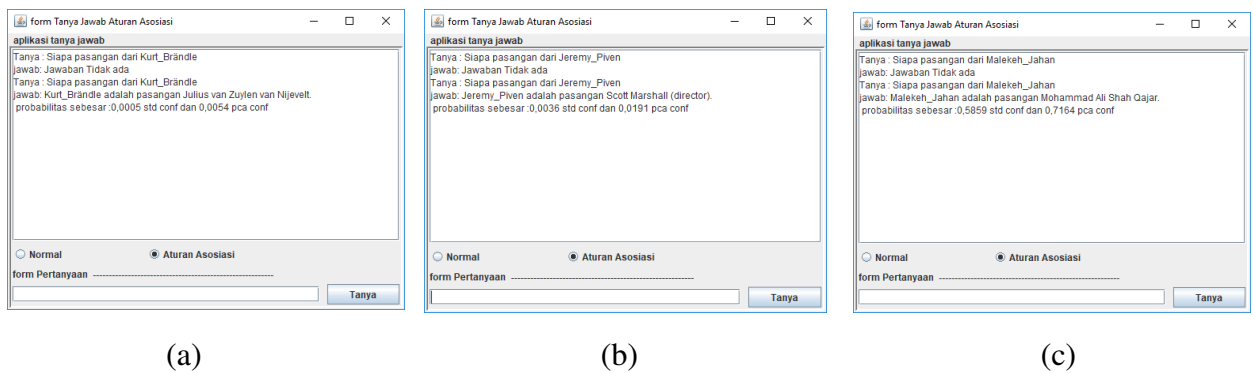

(a)                              (b)                              (c)

Figure 4. Examples of QAGPAR

We tested 7 rules with a standard confidence value of 0.5 and a PCA confidence of 0.5. Our test results showed that 90% of the answers were correct. 10% of the answers were wrong because there was one rule with incorrect logic, namely: 'If B is a citizen from country A, and B graduates from campuses C and D works on campus C, then D is a citizen of A.' For rules that have confidence below 0.5, 90% of the answers given are wrong, while only 10% of the answers are correct, because the rule is logically incorrect, for example: 'If A died in city B, C was born in city B, and C is the leader of country D, then A is a citizen of D.'

## 6. Conclusion

The introduction of huge structured knowledge repositories has opened up opportunities for answering questions, even when the answers are not in a database. In this study, graph-pattern association rules involving multiple relations between entities were used by operating directly on the semantic representations rather than finding answers in natural language. Query languages for accessing these repositories are well-established; however, they are too complicated for non-technical users, who would prefer to pose their questions in natural language. We have presented the QAGPAR system, which gives access to the accumulated knowledge of Wikipedia via the automatically acquired YAGO knowledge base. Our results showed that confidence values above 0.5 will give the correct answer and confidence values below 0.5 will give false results. However, the value of confidence alone is not enough to get the correct answer. Another important factor is the logic of the rules. Rules that have correct logic tend to give right answers, even when the confidence value is low. Rules that have incorrect logic tend to give wrong answers, even if the confidence value is high.

Answering questions by structured knowledge queries', *Proc. - 5th IEEE Int. Conf. Semant. Comput. ICSC 2011*, pp. 158–161, 2011.